%% file: Vela.tex
\documentclass{aa}  

\usepackage{graphicx}
\usepackage{txfonts}
\usepackage{natbib}
\usepackage{hyperref}
\usepackage{physics}
\bibpunct{(}{)}{;}{a}{}{,} 

\usepackage{multirow}

\usepackage{color}

\newcommand\Fixed[1]{}

\newcommand{\kpc}{{\rm kpc}}
\newcommand{\magnitude}{{\rm mag}}
\newcommand{\magkpc}{mag~kpc$^{-1}$}
\newcommand\loneq[1]{\ensuremath{\ell = #1\degr}}

\newcommand\lonbe[2]{\ensuremath{#1\degr \leq\ell \leq #2\degr}}
\newcommand\bonbe[2]{\ensuremath{#1\degr \leq b \leq #2\degr}}
\newcommand\deq[1]{\ensuremath{D= #1~\kpc}}

\newcommand\deqpm[2]{\ensuremath{D= #1\pm#2~\kpc}}

\renewcommand\ao{A_0}
\newcommand\prob[1]{P\left(#1\right)}

\newcommand\probc[2]{P\left(#1 \mid #2\right)}

\newcommand\gbp{G_{BP}}
\newcommand\grp{G_{RP}}

\newcommand\fedred{FEDReD}
\newcommand\lbloc[2]{\ensuremath{(\ell,b) = (#1\degr, #2\degr)}}
\newcommand\pc{\ensuremath{\rm pc}}
\newcommand\toprule{\hline\hline}
\newcommand\bottomrule{\hline\hline}
\newcommand\midrule{\hline}
\renewcommand\deg{\ensuremath{^{\circ}}}

\begin{document}

\title{FEDReD III : Unraveling the 3D structure of Vela}

\author{
  C. Hottier \inst{1} 
  \and 
  C. Babusiaux \inst{2,1}         
  \and
  F. Arenou \inst{1}
}

\institute{
  GEPI, Observatoire de Paris, CNRS, Universit\'e Paris Diderot ; 5 Place Jules Janssen 92190 Meudon, France
  \and 
  Univ. Grenoble Alpes, CNRS, IPAG, 38000 Grenoble, France
}

\date{Received ; accepted }

  \abstract
  {The Vela complex is a region of the sky that gathers several stellar and
  interstellar structures in a few hundred square degrees.}
  {Gaia data now allows us to obtain a 3D view of the Vela interstellar structures through
  the dust extinction.}
  {We used the FEDReD (Field Extinction-Distance Relation Deconvolver) algorithm on
    near-infrared 2MASS data, crossmatched with the Gaia DR2 catalogue, to obtain a 3D cube
    of extinction density.  We applied the FellWalker algorithm on this cube to locate clumps
  and dense structures.}
  {We analysed 18 million stars on $450~\mathrm{deg}^2$ to obtain the extinction
    density of the Vela complex from 0.5 to 8~kpc at $\ell\in[250\degr,280\degr]$ and
    $b\in[$-10$\degr,5\degr]$. This cube reveals the complete morphology of known structures
    and relations between them. In particular, we show that the Vela Molecular Ridge is more
    likely composed by three substructures instead of four, as suggested by the 2D densities.
    These substructures form the shell of a large cavity. This cavity is visually aligned with
    the Vela supernova remnant but located at a larger distance. We provide a catalogue of
    location, distance, size and total dust content of ISM clumps that we extracted from the
  extinction density cube.}
  {}


   \maketitle
%

\section{Introduction}

\begin{figure*}[ht]
  \centering
  \includegraphics{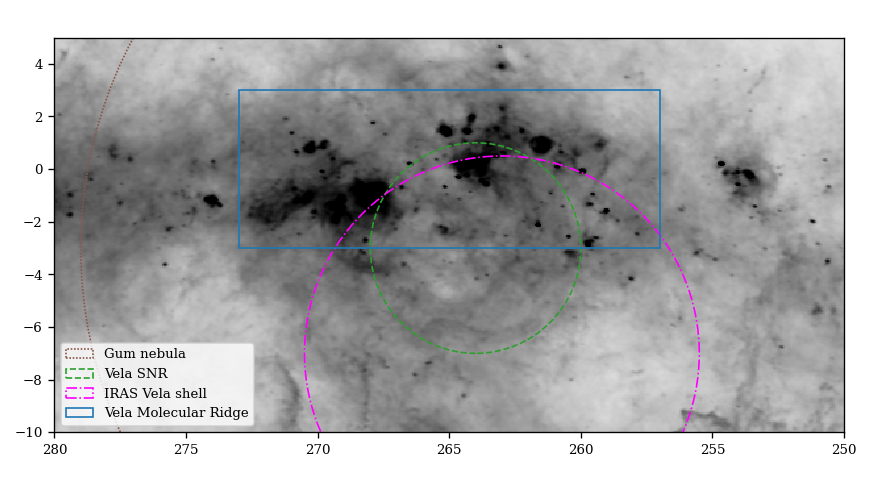}
  \caption{View of the Vela complex at $60~\mu{\rm m}$ with the Improved Reprocessing of the
  IRAS Survey \citep[IRIS]{miville-deschenes2005}. {The colour map is on a logarithmic scale}. We overlay the approximate location of the main Vela substructures discussed in the text.
}%
  \label{fig:IRIS}
\end{figure*}

A large number of structures are known towards the Vela constellation. A view from the Improved
Reprocessing of the IRAS Survey \citep[IRIS]{miville-deschenes2005} of this sky area
is {displayed in} Fig.~\ref{fig:IRIS}.
This is the Vela complex. All of its structures reside in the region defined by \lonbe{240}{280} and \bonbe{6}{-20}
\citep{pettersson2008}. As stellar structures, we count several OB association as well
as R associations \citep{pettersson2008}. However, the Vela complex also presents several large interstellar structures. 

The Gum nebula \citep{gum1952} is the largest structure of the area \citep{pettersson2008}. It
is a ring shell structure with an angular diameter of $34\degr$ centred at \lbloc{262}{-3}\
\citep{reynoso1997} at a distance of $450\pc$. The nebula is expanding and shares this
expansion with the cometary globules \citep{woermann2001} which are a set of dark clouds with
tails pointing to the centre of the Gum nebula \citep{pettersson2008}. 

The Vela surpernova remnant (VSNR) structure is located at \lbloc{263.9}{-3.3}, centred on the Vela Pulsar, PSR B0-833-45, with an angular diameter
of~$\approx8\degr$\citep{pettersson2008}. The estimated distance varies from $250\pc$ to $600\pc$
\citep[and references therein]{cha1999}.

The IRAS Vela shell (hereafter IVS) is a ring shell first noticed in IRAS data by
\cite{sahu1992a}. It is centred at \lbloc{263}{-7} at the distance of $450\pc$
\citep{pettersson2008}. This shell is related to the Vela OB2 association and may have a common
progenitor and may share their history \citep[and references therein]{cantat-gaudin2019a}.

It is canonically accepted that all these structures are quite close to the Sun.  The Vela
complex nevertheless also presents a large ridge in the background. It was first noticed in
\element{CO} Survey \citep{dame1987, may1988} and named Vela Molecular Ridge (VMR) by
\cite{murphy1991}. This structure can be split in four parts, A, B, C and D, according to
\element{CO} peaks. \cite{liseau1992} estimated the distances to VMR A, C and D at
$0.7\pm0.2\kpc$ and VMR B at $2\kpc$, using infrared photometry which could imply that VMR B is
not related to the A, C and D parts.  More recently, \cite{massi2019} used stellar parallax
from Gaia DR2 \citep{gaiacollaboration2018a} to estimate the VMR C distances farther at
\deq{0.95\pm0.05}.

As shown by the previous paragraph, the Vela region has been studied through several markers and
techniques, but the extinction component, especially in 3D, is poorly exploited compared to other tracers.
\cite{franco2012} analyses the interstellar reddening in six distinct areas using photometry to
study the largest structures of the Vela complex except the densest parts of the complex.

On the other hand, there are some 3D extinction maps, but some of them are focused on other
regions, such as \cite{chen2013b} or  \cite{schultheis2014}, which compare photometric surveys
to the Besançon model \citep{robin2012d} to study the Galactic bulge. \cite{sale2014} and
\cite{green2018} both used a Bayesian approach, respectively on IPHAS survey and on Pan-STARRS
and 2MASS, to derive extinctions.  \cite{rezaeikh.2018} analysed APOGEE survey with a
non-parametric 3D inversion to obtain the dust density.   But these three techniques used
northern sky surveys, so they are not able to reach Vela.

There are also some 3D extinction maps which cover the Vela direction. \cite{capitanio2017b} and 
\cite{lallement2019} applied the 3D inversion technique of \cite{vergely2001} on composite dust
proxies using respectively the Gaia DR1 parallaxes and the 2MASS crossmatch with Gaia DR2.
\cite{chen2019a} used a random forest algorithm on a dataset built with Gaia DR2, WISE and 2MASS
to compute the extinction density. \cite{hottier2020} analysed 2MASS and Gaia DR2 data with 
the \fedred\ algorithm \citep{babusiaux2020} and derived the extinction density in the
Galactic plane. 

However, as far as we know, there is no study focussing on the 3D extinction structure of
the Vela complex. In a previous work \citep{hottier2020}, we noticed that the Vela complex
presents large structures that reach distances up to $5$kpc. We also discussed in \cite{hottier2020} the possibility that
the Vela complex could belong to the local arm. This belonging seems to be confirmed by the spiral arms 
locations found by \cite{khoperskov2020}. Studying the Vela complex could therefore give the opportunity
to study a spiral arm viewed from the inside. 

In this work we used \fedred\ to probe the 3D extinction density distribution towards Vela.
In section~\ref{sec:data} we present the analysed dataset. Section~\ref{sec:fedred} sums
up the \fedred\ algorithm and develops the differences from \cite{hottier2020}. In
Sec~\ref{sec:fellwalker} we present the method of clump extraction.
Section~\ref{sec:results} describes and analyses the three dimensional density map and the clumps and cavities extracted from it.

\section{Data}
\label{sec:data}
In this study, we use the infrared photometry in bands $J$, $H$ and $K$ from the 2MASS
survey \citep{skrutskie2006} and we combine them with the photometry in bands $G$, $G_{BP}$ and
$G_{RP}$ and the astrometry both coming from the Gaia DR2 \citep{gaiacollaboration2018a}.
We use the same methodology as \cite{hottier2020} to filter and merge data of these surveys, so
we will just briefly review them here.

We use the 2MASS near-infrared photometry as
principal data, that is our dataset completeness will be the 2MASS one.
In practice, every star included in our dataset has \texttt{ph\_qual}$\ge$D
in all three $J$, $H$ and $K$ bands. Once we have selected the stars in 2MASS, we use the
\cite{marrese2019} crossmatch to potentially add Gaia DR2 parallaxes and photometry.

To filter the Gaia photometry \citep{evans2018}, we do not use $\gbp$ and $\grp$ when
\texttt{phot\_bp\_rp\_excess\_factor} $> 1.3+0.06 \times(\gbp-\grp)^2$ and $\gbp>18$ according
to \cite{evans2018, arenou2018a}. For the astrometric information \citep{lindegren2018}, we use
the equation 1 of \cite{arenou2018a}, we correct the parallax zero point of $-0.03$~mas and we
do not use the parallax when $\varpi + 3 \times \sigma_{\varpi}< 0$ to remove 
spurious astrometric solutions.

\section{Extinction map with \fedred}%
\label{sec:fedred}

To analyse this  dataset, we used the \fedred\ algorithm. The entire algorithm description as
well as tests on mock and observed data are presented in \cite{babusiaux2020}. Thus we will
just briefly explain the main steps of the algorithm and differences from \cite{hottier2020}.

\subsection{Field of view analysis}%
\label{sub:fiel_of_view_analysis}

\fedred\ analyses photometric and astrometric data, field of view by field of view, and
infers both the evolution of the extinction as a function of distance and the stellar density distribution. To do so, it works in two steps.

The first step is the processing for each star of the likelihood of this observed star being at
the distance $D$ with extinction $\ao$ (extinction at $550$nm) $\probc{O}{\ao,D}$. This
distribution is processed by comparing the apparent photometry of the stars to an empirical HR
diagram built from 2MASS and Gaia DR2 \citep[see][section 3.1 for details]{babusiaux2020}, it
also uses parallax information when it is available.

Once the likelihood of each star computed, \fedred\ applies a Bayesian deconvolution to
obtain the joint distribution of extinction and distances $\prob{\ao,D}$. This
deconvolution also takes into account the completeness of the field of view, which is estimated from the
observed near-infrared photometry distribution. 

To initialise this deconvolution, we use two simple priors.
The prior on the distance distribution is a square law of the distance, corresponding to
 the cone effect. Concerning the extinction given the distances, $P\left(\ao \mid D\right)$, we use an
uniform distribution \citep[contrary to][]{hottier2020}.

From the joint distribution of extinction and distance, \fedred\ generates Monte-Carlo Solutions
(MCSs)
of the increasing relation $\ao(D)$ drawn following the probability distribution $\prob{\ao,D}$.
As Red Clump stars are the ones {providing the strongest constrains on }the distance/extinction, 
we restricted our results to the distance interval $[D_{\rm min}, D_{\rm max}]$ where red clump stars are observed by 2MASS,
that is the distance at which a red clump star saturates ($D_{\rm min}$) or is fainter than the completeness limit ($D_{\rm max}$).
Unlike \cite{hottier2020}, this distance interval restriction is also used within the algorithm determining the $\ao(D)$ and not just 
in the post-processing. 
At the end of this process we obtain $1000$ MCSs by field of view. 

\subsection{Merging fields of view into extinction cubes}%
\label{sub:merging_fields_of_view_into_maps}

The fields of view are $0.38\deg$ wide in longitude and latitude, and the centers are spaced by
$0.38\deg$ in latitude and longitude. Therefore each field of view overlaps its first neighbour
(top, bottom, left and right) by half of its angular surface and overlaps its second neighbour
(four diagonals) by the quarter of its angular surface. This means that each star
is inside three different field of view, which allows us a good continuity between fields of
view.

To merge results from each field of view into a consistent extinction cube, we use the exact same
algorithm as in \cite{hottier2020}. Firstly, we iteratively clean the MCS samples of each field of view
by using neighbour fields' MCS envelopes as upper and lower limits, the convergence of this
process being provided by the overlapping of fields. Secondly, we randomly draw one MCS per field of view
(inside their respective clean pool) to obtain a relation between the extinction $A_0$ and the
distance $D$ for each field of view. 
Then, at each distance bin we smooth the extinction value using the eight neighbour fields of view, to obtain the
extinction cube.
We randomly draw $100$ of these cubes. Finally, we use a constrained
cubic spline fit \citep{ng2007} to obtain the median relation of extinction as a function of
distance of each field of view. These relations are then decumulated and normalised by the
distance width of bins to obtain the extinction density $a_0$ in each voxel of the cube. 

\subsection{Extinction uncertainty}%
\label{sub:density_uncertainty}

As in \cite{hottier2020}, we also compute the uncertainty of our extinction and 
extinction density cubes. Concerning the extinction $A_0$ we use the sample of MCSs after the
cleaning process by the neighbour envelope to get the maximum and minimum values of extinction
at each distance bin (${A_0}_{\rm min}(D)$ and ${A_0}_{\rm max}(D)$). 

To estimate the extinction density uncertainty, we use the exact same algorithm explained in
section \ref{sub:merging_fields_of_view_into_maps} but on a bootstrapped MCS sample instead. We
build $100$ bootstrap merged cubes and we compute the standard deviation of extinction
density $a_0$ for each voxel to get the uncertainty on the extinction density. As
discussed \cite{hottier2020} (section 5.1), this uncertainty map mostly represents the
sampling error, and it underestimates the true uncertainty of our results. 

\section{Clump Extraction}%
\label{sec:fellwalker}
To obtain distance, shape and size of extinction density clumps, we interpolate our data cube
on a regular Galactic Cartesian grid\footnote{Using interpolate.interpn from the SciPy library
with the linear mode}
with a voxel size of 10 pc.

Using this new interpolated cube, we first tried to process the iso-surface density to locate clumps.
This allowed us to spatially constrain some clumps but it required very sensitive settings for
each clump we looked for. Moreover, this technique has difficulty resolving 
distinct clumps, and local peaks can hide the true shape and size of a clump. 

Dendograms have already been used to extract clumps and molecular clouds
\citep{goodman2009, chen2019a}. This algorithm uses hierarchical tree to segregate structures,
showing relations between each clump. To do so it only uses the local value of pixels and
links voxels with their neighbours if values are compatibles. But as we work with a density cube
processed by line of sight, the extinction density can leak to larger distances and create some
fingers of god structures.

To avoid the extraction of structures only dominated by fingers of god effect, we use the FellWalker algorithm\footnote{We re-implemented the algorithm from
scratch in pure python in order to deal with spherical and cartesian coordinates during the
filtering steps.} \citep{berry2015} which identifies clumps by
analysing the local gradient. In a few words, this algorithm "walks" through the entire volume,
voxel by voxel. To choose the next voxel it always chooses the steepest path. When it reaches
a local maximum, it checks nearby for a higher point (in order to avoid extrema due to
noise). If
  this higher point exists the algorithm "jumps" and so on. If a higher spot does not exist, the
algorithm has just reached the top of a new clump. When the entire volume is mapped the algorithm check if
clumps can be merged. Finally, it can reassign a voxel to the most common clump around it.

As the finger of god effect also impact the gradient of extinction density, the
FellWalker algorithm is also affected. While a minimum value based algorithm (such as dendogram) just
wipe out low extinction structures to avoid elongations, 
the FellWalker algorithm ability of setting up both a minimal value and a
minimum gradient can mitigate the extraction of finger of god effect when looking for the edge
of clumps, without missing the ones with low values of extinction.
Nevertheless, the clumps segregation is 
depending on the parameter settings of the algorithm.

  With this in mind, we manually tune  the parameters of the algorithm in order to avoid elongations
  and extract parts of the VMR.
The
minimum extinction density to be a part of a clump is $3$~\magkpc. The size of the
maximum jump is set to $4$ voxels which corresponds to $0.04~\kpc$. The minimum initial density variation is
$0.5$~\magkpc\ from one voxel to another. 
Two clumps may be merged if the depth of the col between them is
smaller than $3$~\magkpc. The clump outline process is
done with a cube of $4$ voxel. Finally, we removed clumps which contain only one or two voxels as 
those are just high extinction density spots induced by the decumulation process.

\subsection{Processing clump parameters}%
\label{sub:processing_clump_parameter}

The result of the FellWalker algorithm is a $3$D mask for each clump. From this mask we
directly obtain angular and distance bound of clumps as well as the volumes of the clumps. 
To obtain the center of a clump, we compute the barycentric position
using extinction density as weight.

In order to estimate the total dust content $A_c$ of a clump we have to integrate the extinction density
$a_0$ over the volume of the clump:
\begin{eqnarray}
  A_c &=& \int_{\Omega}  a_0 \dd{V},\label{eq:AC1}\\
  A_c &=&  \int_{\Omega}  a_0 d^2 \cos{b} \dd{l} \dd{b} \dd{d}, 
\end{eqnarray}
if we discretise the equation we get: 
\begin{equation}
  A_c = \sum_{i\in\Omega} {a_0}_i\delta d_i\ {d_i}^2 \cos{b_i} \delta l_i \delta b_i, 
\end{equation}
with $i$ the $i^{th}$ voxel in the clumps, $l_i$, $b_i$ and $d_i$ the position of the voxel,
${a_0}_i$ the extinction density in this voxel and $\delta d_i$ the distance width of the
voxel. As these voxels follow the cube we obtained by merging the fields of view, $\delta l_i$ and
$\delta b_i$ are constant over the cube $\delta l=\delta b=0.18\deg$. 
In practice, as we looked for clumps in a regular cartesian cube, we used the
Eq~\ref{eq:AC1} in cartesian referential frame with $\dd{V}=(0.01\kpc)^3$ and $a_0$ the value
of extinction density in each voxel.
Note that these $A_c$ can be used to estimate the masses of clumps, following the method described in \cite{chen2020}.

Some of the clumps are located on the edge of our extinction cube. We mark them as "not full" and
their estimated extinction is only a lower estimate while their distance only corresponds
to what we observe in our extinction cube.

\subsection{Uncertainties on the parameters}%
\label{sub:uncertainties_of_parameters}

To estimate the uncertainty on each parameter, we used the random cubes that
we have built to compute the density uncertainty. We apply on them the same FellWalker algorithm
with the same parameters. We cross identify clumps from their barycentric position and we
estimate each parameter uncertainty using their standard deviation.

We kept, in the final catalog, clumps which are detected in at least $99\%$ of the
bootstraps. Indeed, some large clumps are identified as several small ones in some bootstraps,
preventing the usage of a 100\% threshold.

\subsection{Extracting cavities}%
\label{sub:extracting_cavities}

We also used the FellWalker algorithm to spatially constrain density cavities. As this
algorithm is designed to search peaks rather than valleys, we inverted the
values of the density cube following this equation : $\max(a_0(l,b,d))- a_0(l,b,d)$. The
cavities in the "inverted" data cube appear more as a plateau than as a mountain so we did not use
the minimum steepness parameter. The minimum value parameter is set in order to map the area of the
original cube with extinction less than $0.5~\magnitude$. As for the clump extraction we set the
maximum jump to $4$ voxels.

Inverting the data cube implies that every area without extinction is detected as a cavity, so
areas above and below the complex are considered as many little cavities which merge into a
single one during
the merging process.
Moreover, the shell of cavities that we can see by eye are porous, which leads them to be merged
with the clean empty part below and above the complex. To avoid this phenomenon, we clean our
cavity sample before the merging step. To do so we remove from the sample the cavities with a
median voxel values inferior to $0.1~\magnitude.\kpc^{-1}$. Then we merge the cavities with a col
height threshold at $0.03~\magnitude.\kpc^{-1}$.

  Finally, we remove cavities on the edge of our field of view. Indeed, in the case of
  clumps, if a structure is detected on the edge it is for sure a clump.
  But in the case of cavity detection, we are looking for area with low extinction density. In
  the case of cavity inside our data cube we can ensure that this area is really surrounded by
  an extinction shell, while a cavity on the edge of the cube can be completely open.

As for clump, we also processed uncertainty on cavities parameters, but due to the large volume
of the cavities and the thin shell around them, the tuning of the parameters is very
sensitive. This implies that we do not recover cavities in bootstrap as often as we did for
clumps. So, in order to provide uncertainties, we lower the threshold at $50\%$ to keep
cavities in final catalog.

\section{Results}%
\label{sec:results}
\begin{figure*}[!h]
  \centering
  \includegraphics{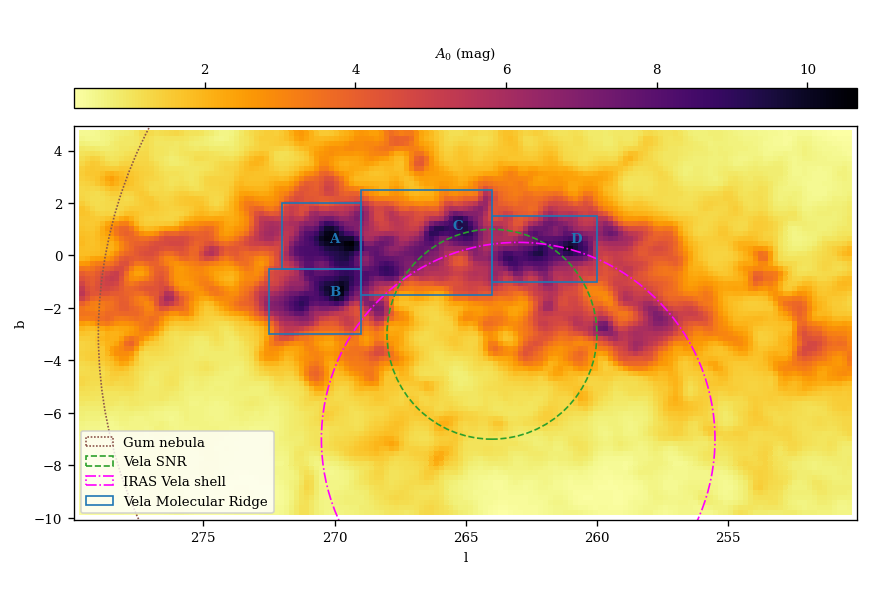}
  \caption{View of the Vela complex in extinction $\ao$ using the cumulative extinction over a distance of $5~\kpc$.}%
  \label{fig:fullview}
\end{figure*}

\begin{figure*}
  \centering
  \includegraphics{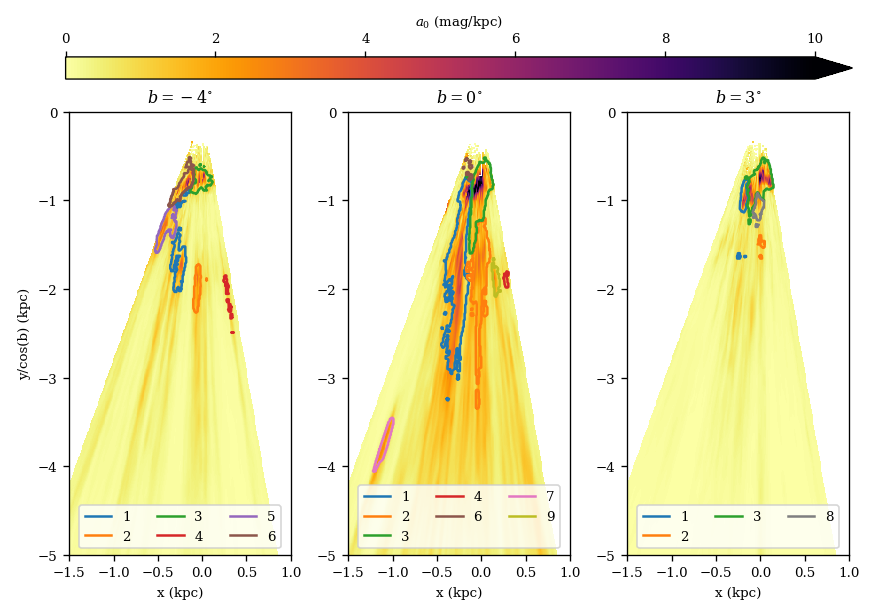}
  \caption{Extinction density $a_0$ at different Galactic latitudes $b$. $x$ and $y$ represent
  galactic coordinates, the Sun is at $(0,0)$ and the Galactic centre
direction is to the right. We also add the contour of the clumps, their parameters being
presented in Table~\ref{tab:clumps}.}%
  \label{fig:lcut}
\end{figure*}

\begin{figure*}
  \centering
  \includegraphics{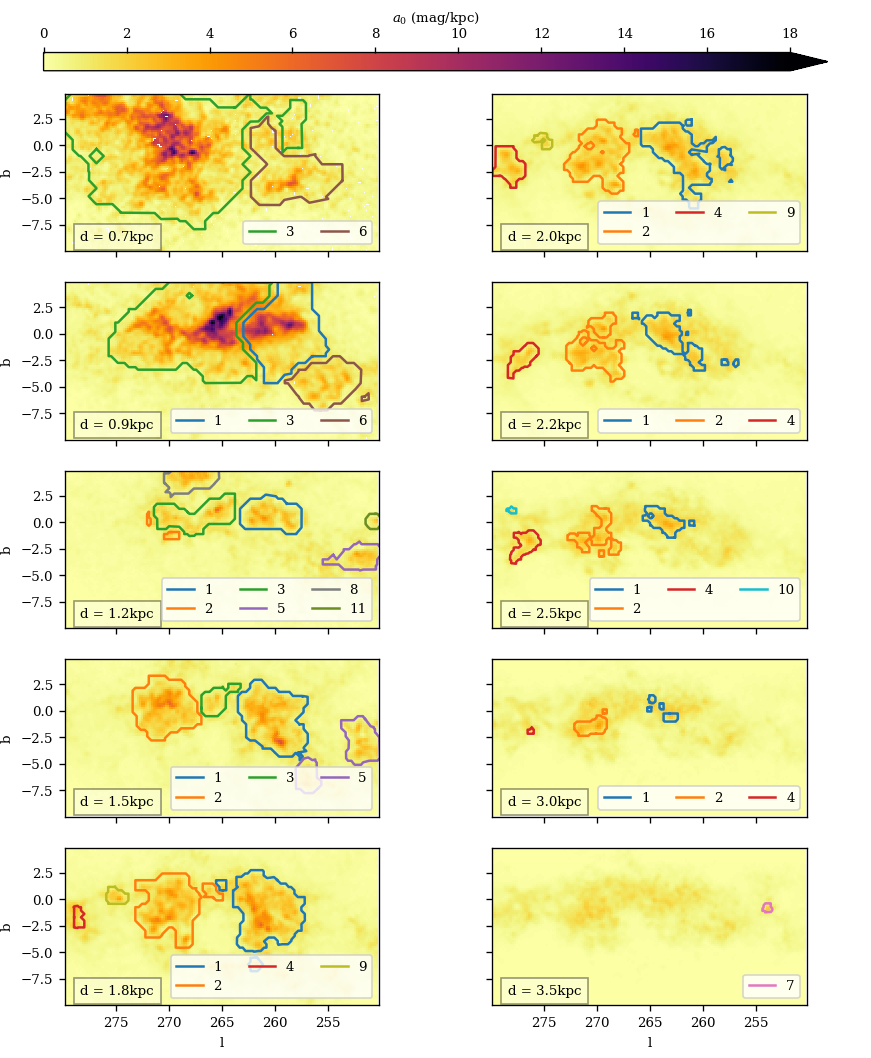}
  \caption{Extinction density $a_0$ of the Vela complex at several distances obtained with
  \fedred. The contours of each clump segmented by the FellWalker algorithm is represented with
lines.}%
  \label{fig:fullviewcut}
\end{figure*}

\begin{figure*}
  \centering
  \includegraphics{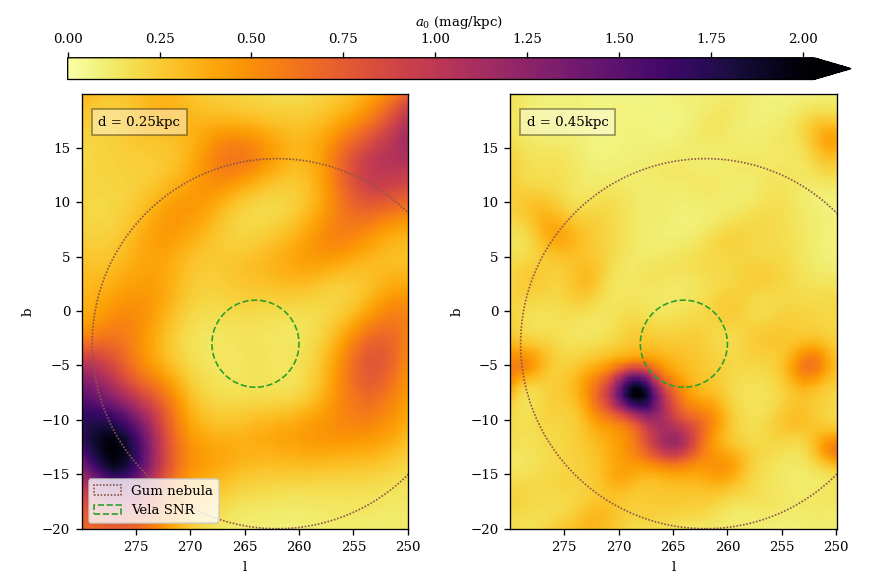}
  \caption{Extinction density of the Vela complex at the distance \deq{0.25} and \deq{0.45} obtained with the
    \cite{lallement2019} cube. The canonial size and location of Vela SNR and Gum Nebula
    \citep{pettersson2008} are
sketched.}
  \label{fig:gumnebula}
\end{figure*}

We split our dataset of $18\,028\,023$ stars into 12\,403 fields of view. Each field of view is a square of
$0.38\degr$ in Galactic latitudes and longitudes, with $l\in[250\degr, 280\degr]$ and
$b\in[-10\degr,5\degr]$.

\subsection{Extinction column}%
\label{sub:extinction_column}
We present in Fig.~\ref{fig:fullview} a view of the 
cumulative extinction $A_0$ of the Vela complex up to $5$~kpc, roughly the distance limit of Vela
complex noticed in \cite{hottier2020}, with the same boundaries as in Fig.~\ref{fig:IRIS}. We also
overplotted the main structures of Vela.

The VMR appears as the strongest extinction structure of the entire complex and is approximately
located at \lonbe{260}{272} and \bonbe{-4}{3}. While the overall shape of the VMR is quite similar
to the IRAS view shown in Fig.~\ref{fig:IRIS}, it differs in the details, as the strongest
emissions at $60~\mu{\rm m}$ do not coincide with the areas with the strongest
cumulative extinction.
As in Fig.~\ref{fig:IRIS}, the IVS is very faint but using a visual guide, it is still visible as a diffuse filament. 
On the other hand, due to the relatively small area covered by our study, the Gum nebula
footprint cannot be distinguished.

\subsection{Extinction density towards Vela}%
\label{sub:vela_and_puppis_in_extinction}

In Fig.~\ref{fig:lcut} we present the extinction density $a_0$ cut along constant Galactic
latitudes.  Fig.~\ref{fig:fullviewcut} shows views of the extinction density at several
distances. We also add on those two figures, the contours of clumps that we segmented with the
FellWalker algorithm. 

On both figures, we do not provide the extinction density close to the Sun because of
the red clump visibility criterion (see Sect~\ref{sub:fiel_of_view_analysis}). 
To see the closeby Gum Nebula and the VSNR, we used instead the \cite{lallement2019} extinction cube. They used the
same data as this work (2MASS and Gaia DR2) restricted to stars with relative error on the
parallax less than $20\%$, and they apply the same extinction law (see 
Appendix.~\ref{sec:comparison_with_lallement2019} for a more detailed comparison). In Fig.~\ref{fig:gumnebula}
we plot the extinction density that they obtain at the canonical distance of the Gum Nebula
and VSNR. Indeed, at \deq{0.25} we saw a shell which has the size of the VSNR. We also see
a large shell which seems to correspond to the Gum Nebula at \deq{0.25} whereas we cannot
distinguish any footprint at \deq{0.45}. This could mean that the Gum Nebula is closer than 
usually indicated in the literature.

\subsection{VMR and Clumps}

\begin{sidewaystable*}
  \centering
  \caption{Density clumps found by the FellWalker algorithm. The coordinates are in the Galactic
  referential centred on the Sun and X being towards the Sun. 
  \label{tab:clumps}}
  \input{figures/clumpTable_2.tex}
\end{sidewaystable*}

\newcommand\clumpnum{$14$}
\newcommand\vmrcid{$3$}
\newcommand\vmraid{$2$}
\newcommand\vmrdid{$1$}
\newcommand\topclumpid{$8$}
\newcommand\rcwdnid{$7$}

The FellWalker algorithm found \clumpnum\  clumps. Table~\ref{tab:clumps} presents their parameters. We
identify them by a simple integer id sorted by their volumes. We visually identify clumps to the VMR parts (column
Notes of Table~\ref{tab:clumps}).

The VMR C (clump~\vmrcid) goes beyond the edge of our extinction cube but we can see that we detect the main part of it. It
begins at \deq{0.5} and its barycenter is at \deqpm{0.9}{0.09}, which is consistent with the
\cite{liseau1992} estimation at $0.7\pm0.2~\kpc$. This distance is also inside
the confidence interval of \cite{massi2019}($0.950\pm0.050~\kpc$) and our distance interval
contains every molecular distances of \cite{zucker2020} labeled VMRC, which range between 0.86 and 0.97~\kpc. However VMR C is very large, its
longitude footprint goes from \loneq{256} to more than \loneq{280} and its maximum distance
reaches \deq{1.6}.

VMR D (clump~\vmrdid), the biggest of this study, is farther than VMR C. Its barycenter is at \deqpm{1.9}{0.4} 
and its front part is at \deq{0.75}, while \cite{liseau1992} only see the front part of it and estimate VMR C,D and A to be roughtly at the same distance of $0.7\pm0.2~\kpc$.

Concerning VMR A and VMR B, they are identified as a unique clump
(clump~\vmraid)\footnote{Clump~\vmraid\ could by segmented into two distinct clumps but only
with stronger settings of the FellWalker algorithm}. 
The reason why this clump is split into two
pieces in the literature, is the overlapping with VMR C in the foreground which induces a projection effect.
Nevertheless, VMR A is one of the biggest clump of the Vela region and it seems to be related to
VMR C and VMR D at high distances (${\rm D} > 1.2 \kpc$) where they form a shell (see
Sect~\ref{sub:cavities}). We can also note than this clump contains the molecular
cloud RCW~38 which is located at \deq{1.6} by \cite{zucker2020}.

Clump~\rcwdnid\ corresponds to the RCW~19 object \citep{rodgers1960} also called GUM~10
\citep{gum1955}. We locate it at \deqpm{3.8}{0.09}, which is farther than the determination \deqpm{3.0}{0.3} of \citep{russeil2003}.

\newcommand\closeclumpid{$6$}
As far as we know, the other clumps have not yet been identified and named. 
Clump~\topclumpid\ is a large clump, and it is bigger than what we process  as it goes beyond the edge of our extinction cube.
Clump~\closeclumpid\ presents a relatively high extinction density and reasonable
apparent angle but it is barely visible
in the column view (Fig.~\ref{fig:fullview}) because it is overlapped by VMR~D.

\subsection{Cavities}%
\label{sub:cavities}

\begin{table*}
  \centering
  \caption{Cavities found by the FellWalker algorithm. The X,Y,Z,l,b,d are the coordinates of
    the barycenter of the cavities express  in the Galactic
  referential centred on the Sun.}
  \label{tab:cavities}
  \input{figures/cavitiesTable.tex}
\end{table*}

\begin{figure}[ht]
  \centering
  \includegraphics{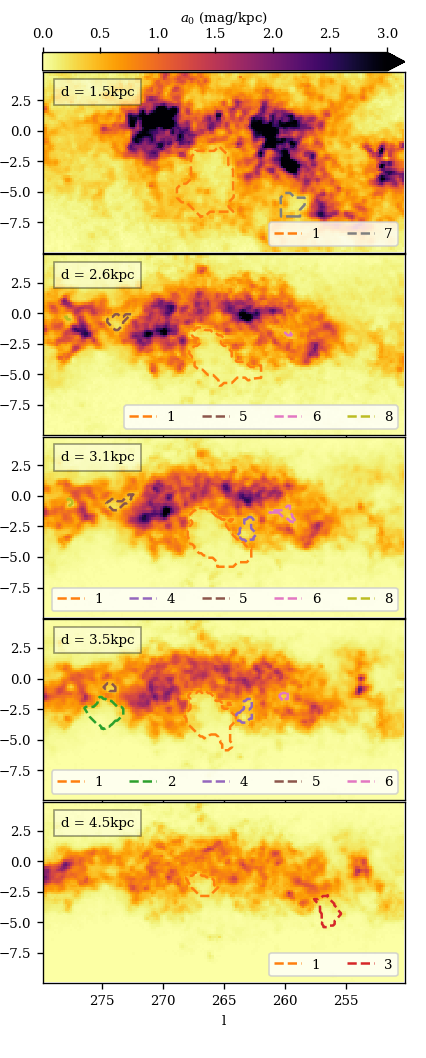}
  \caption{Extinction density $a_0$ of the Vela complex at several distances obtained with
  \fedred. The lines represent the contours of the cavities we extracted with the FellWalker
algorithm (parameters on Tab~\ref{tab:cavities})}
  \label{fig:cavities}
\end{figure}

\newcommand\nbrcav{$9$}
\newcommand\bigcavid{$1$}

The FellWalker algorithm found \nbrcav\ cavities. As for clumps, we identify them by a simple integer ID sorted by
volume. We represent the contour of this cavities in distance views on
Figure~\ref{fig:cavities} and their parameters are presented in Table~\ref{tab:cavities}.

The biggest one (cavity~\bigcavid) is centered on \lbloc{266.1}{-3.1}. It is very visible on IRAS data
(Fig.~\ref{fig:IRIS}) and on the cumulated extinction view (Fig.~\ref{fig:fullview}). Its 
angular location fits the VSNR angular location but the center of this cavity is at
\deqpm{3.1}{0.35} and
its closest "wall" is at \deq{1.2}, behind the foreground part of VMR C (clump~\vmrcid) and its
shell is composed by all the VMR parts. On the other hand, VSNR is at $0.25$~kpc with an upper
limit at $0.49$~kpc \citep{cha1999}, in the foreground of the VMR. We saw its local extinction fingerprint
in Fig~\ref{fig:gumnebula}. So, it appears that this shell, described in the 2D views as VSNR, visible in infrared and in
extinction column, is actually composed of two distinct structures physically separated by VMR C.

As Cavity $1$, Cavities $2$ and $3$ are under the main extinction density of the galactic plane
and their
low latitudes shells are very thin. While the others are located inside the galactic plane. 
So far, we have not found the origins of these cavities. 

\section{Conclusion}%
\label{sec:discussion}

We used the \fedred\ algorithm to analyse photometry and parallax from 2MASS and Gaia DR2
of stars in the Vela complex direction. This provided us with the distribution of the extinction
density in 3D. Consequently we were able to obtain the distance and the 3D shape of known
structures of this area.

For this purpose we applied the FellWalker algorithm, which produced very good results when applied to
extracting clumps. However it turns out not to be the best designed tool to extract cavities
as it requires a very specific tuning. For further investigation on cavities, we recommend
using or developing other approaches, although we have not found a better one so far. 

Nevertheless, we managed to spatially measure \clumpnum dense extinction
clouds (Table~\ref{tab:clumps}) and \nbrcav\ cavities (Table~\ref{tab:cavities}). We then
visually identified these clouds using the projections of the known structures. Doing this, we determined the
distance of Vela Molecular Ridge components. 
It appears that the split of VMR into 4 parts described in the literature is more a construction due to a face-on view
of very large structures, where the foreground is composed by VMR C and VMR D. At high
longitude (\textit{i.e.} the western side) the so called VMR A and VMR B are actually one clump only and prolongate a large
shell completed with the background part of VMR C and VMR D.
Moreover, it came out that what seems to be the infrared shell of the VSNR is
actually mainly due to a background cavity surrounded by the components of the VMR.

This study of the Vela complex in extinction reveals that this area contains structures of
different types and in interaction. In fact this complex is probably a part of
the local arm \citep{hou2014, hottier2020, khoperskov2020}. Therefore Vela is a perfect lab to study a spiral arm from the inside.

Tables~\ref{tab:clumps} and \ref{tab:cavities} as well as data cubes of extinction density, 
the uncertainty of the extinction density, the
cumulated extinction and the minimum/maximum extinction values will be available in a machine-readable from
at the CDS. 

\begin{acknowledgements}
The authors thank the referee for comments and suggestions that improved the paper.
C.H. thanks the Centre National d'Études Spatiales (CNES) for the Gaia post-doctoral grant. 
This work has made use of data from the European
Space Agency (ESA) mission Gaia (https://www.cosmos.esa.int/gaia), processed by the Gaia Data Processing and Analysis Consortium (DPAC,
https://www.cosmos.esa.int/web/gaia/dpac/consortium). Funding for the DPAC
has been provided by national institutions, in particular the institutions participating in the Gaia Multilateral Agreement. 
This work makes use of data 
products from the 2MASS, which is a joint project of the University of Massachusetts and the Infrared Processing and Analysis Center/California Institute
of Technology, funded by the National Aeronautics and Space Administration
and the National Science Foundation.
\end{acknowledgements}

\bibliographystyle{aa} 
\bibliography{FEDReD_Vela}

\appendix
\section{Comparison with \cite{lallement2019}}%
\label{sec:comparison_with_lallement2019}
As we explained in Sec.~\ref{sub:vela_and_puppis_in_extinction}, \cite{lallement2019} use the same
data and extinction law as we do. But their method is very different. In order to
compare the two results we present in Fig.~\ref{fig:stillismcomp00} and
Fig.~\ref{fig:stillismcomp01} extinction density views at different distances produced by both
methods.

We can clearly see that density views of \cite{lallement2019} are smoother than what we get
with \fedred. This is due to their hierarchical inversion with a correlation Gaussian kernel,
which smoothes the structures by fitting a 3D Gaussian at each space location in a cartesian
grid. For this reason the angular resolution is smaller at low distances, while the angular
resolution of \fedred\ is constant and allows us to detect smaller angular structures.

\begin{figure*}
  \centering
  \includegraphics{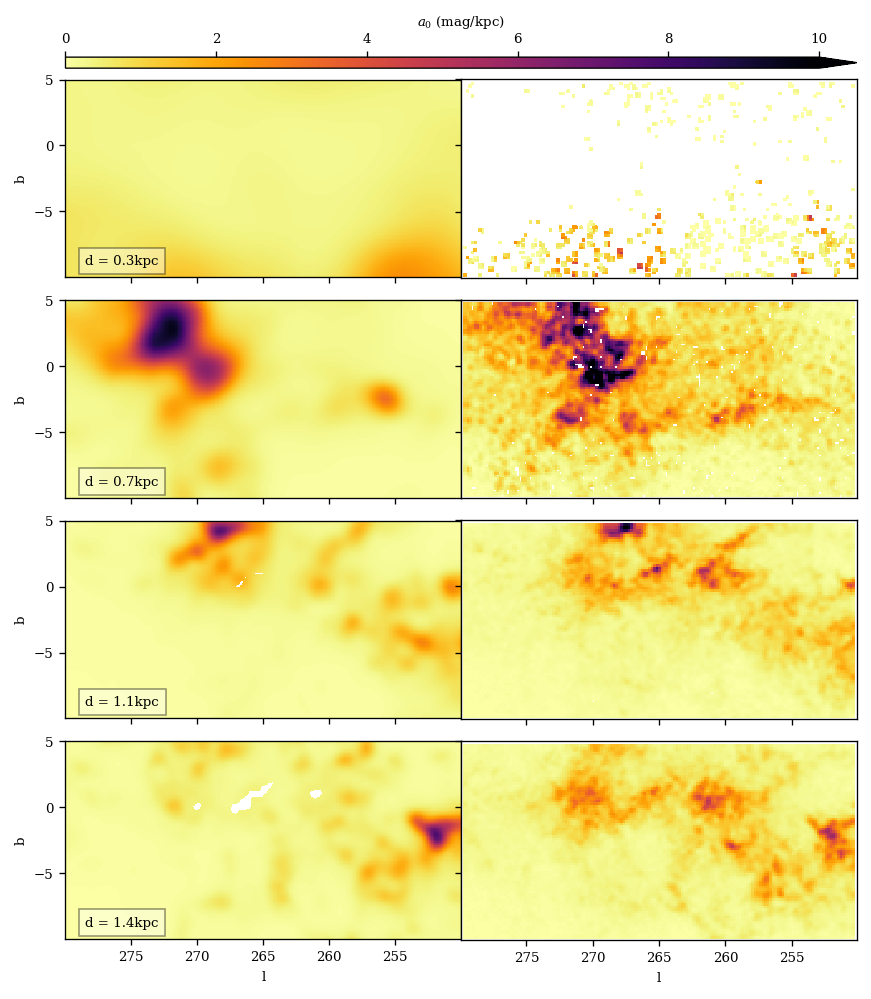}
  \caption{Extinction density $a_0$ by \cite{lallement2019} (left
  panels) and \fedred\ (right panels) at several distances. The \cite{lallement2019} data cube
is interpolated in order to obtain the same resolution as this study.}%
\label{fig:stillismcomp00}
\end{figure*}
\begin{figure*}
  \centering
  \includegraphics{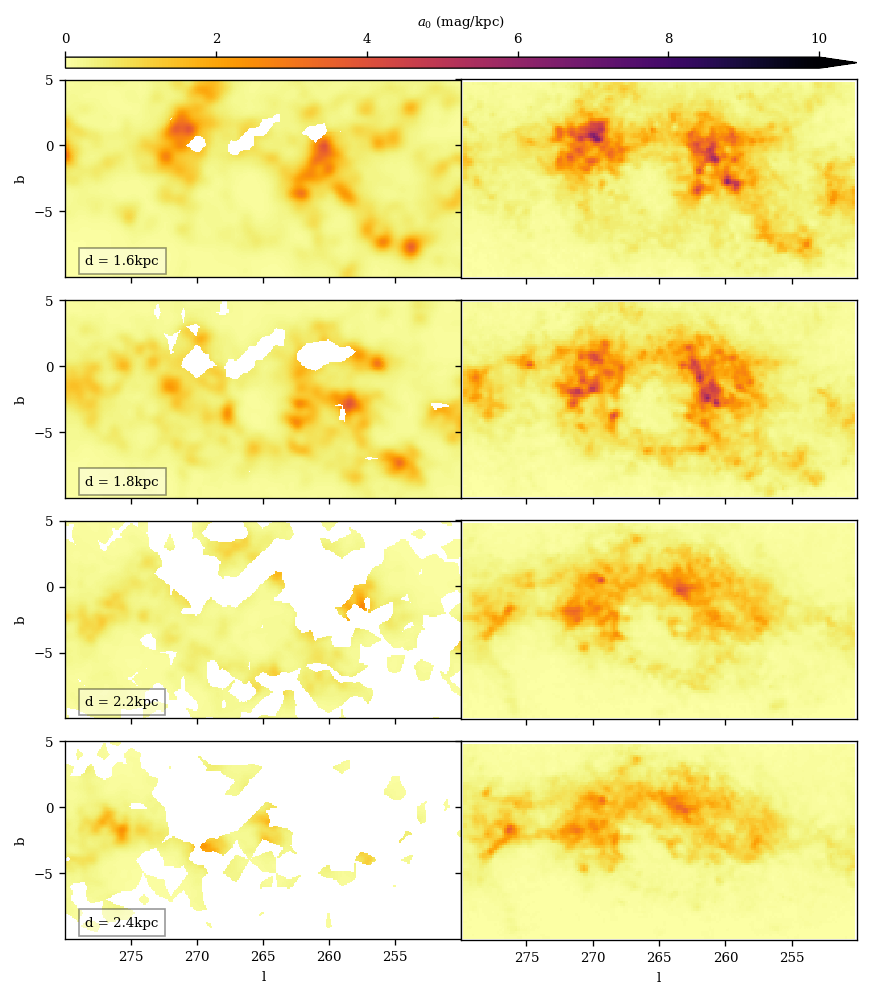}
  \caption{Continuation of Fig.~\ref{fig:stillismcomp00}.}%
\label{fig:stillismcomp01}
\end{figure*}

\end{document}

%% file: figures/clumpTable_2.tex
\begin{tabular}{cccccccccccccccc}
\toprule
\multirow{2}{*}{ID} & X & Y & Z & l & b & d & mind & maxd & minb & maxb & minl & maxl & Volume & AC & \multirow{2}{*}{Notes} \\
 & (kpc) & (kpc) & (kpc) & (deg) & (deg) & (kpc) & (kpc) & (kpc) & (deg) & (deg) & (deg) & (deg) & (pc$^3$) & (mag.pc$^2$) &  \\
\midrule
\multirow{2}{*}{1} &-0.27 & -1.88 & -0.02 & 261.74 & -0.57 & 1.90 & 0.75 & 3.30 & -7.02 & 5.34 & 253.32 & 266.98 & 5.08e+07 & 9.18e+04 & \multirow{2}{*}{VMR D} \\
 &$\pm$ 0.04 & $\pm$ 0.38 & $\pm$ 0.01 & $\pm$ 0.85 & $\pm$ 0.52 & $\pm$ 0.38 & $\pm$ 0.02 & $\pm$ 0.87 & $\pm$ 0.78 & $\pm$ 0.03 & $\pm$ 0.40 & $\pm$ 1.38 & $\pm$ 1.91e+07 & $\pm$ 3.42e+04 & \\
 & & & & & & & & & & & & & & &  \\
\multirow{2}{*}{2} &0.00 & -2.17 & -0.03 & 270.03 & -0.91 & 2.17 & 1.13 & 3.37 & -5.45 & 3.38 & 264.46 & 273.88 & 4.69e+07 & 8.15e+04 & \multirow{2}{*}{VMR A-B} \\
 &$\pm$ 0.00 & $\pm$ 0.01 & $\pm$ 0.00 & $\pm$ 0.02 & $\pm$ 0.02 & $\pm$ 0.01 & $\pm$ 0.06 & $\pm$ 0.07 & $\pm$ 0.04 & $\pm$ 0.44 & $\pm$ 0.43 & $\pm$ 0.24 & $\pm$ 6.74e+05 & $\pm$ 9.13e+02 & \\
 & & & & & & & & & & & & & & &  \\
\multirow{2}{*}{3} &-0.03 & -0.90 & 0.01 & 268.21 & 0.52 & 0.90 & 0.47 & 1.66 & -7.98 & 5.54 & 256.61 & 280.54 & 1.64e+07 & 3.31e+04 & \multirow{2}{*}{VMR C, not full} \\
 &$\pm$ 0.00 & $\pm$ 0.09 & $\pm$ 0.00 & $\pm$ 0.18 & $\pm$ 0.13 & $\pm$ 0.09 & $\pm$ 0.02 & $\pm$ 0.19 & $\pm$ 0.22 & $\pm$ 0.03 & $\pm$ 1.22 & $\pm$ 0.05 & $\pm$ 4.64e+06 & $\pm$ 8.06e+03 & \\
 & & & & & & & & & & & & & & &  \\
\multirow{2}{*}{4} &0.29 & -2.29 & -0.08 & 277.18 & -1.86 & 2.31 & 1.63 & 3.06 & -4.35 & 0.21 & 275.19 & 280.04 & 1.09e+07 & 1.53e+04 & \multirow{2}{*}{not full} \\
 &$\pm$ 0.00 & $\pm$ 0.01 & $\pm$ 0.00 & $\pm$ 0.02 & $\pm$ 0.02 & $\pm$ 0.01 & $\pm$ 0.02 & $\pm$ 0.01 & $\pm$ 0.03 & $\pm$ 0.00 & $\pm$ 0.02 & $\pm$ 0.01 & $\pm$ 1.41e+05 & $\pm$ 1.77e+02 & \\
 & & & & & & & & & & & & & & &  \\
\multirow{2}{*}{5} &-0.40 & -1.35 & -0.10 & 253.42 & -3.94 & 1.41 & 1.11 & 1.74 & -8.56 & -0.14 & 249.83 & 258.33 & 6.60e+06 & 9.02e+03 & \multirow{2}{*}{not full} \\
 &$\pm$ 0.00 & $\pm$ 0.01 & $\pm$ 0.00 & $\pm$ 0.03 & $\pm$ 0.09 & $\pm$ 0.01 & $\pm$ 0.01 & $\pm$ 0.03 & $\pm$ 0.27 & $\pm$ 0.20 & $\pm$ 0.01 & $\pm$ 0.11 & $\pm$ 1.63e+05 & $\pm$ 1.96e+02 & \\
 & & & & & & & & & & & & & & &  \\
\multirow{2}{*}{6} &-0.22 & -0.83 & -0.06 & 255.42 & -3.86 & 0.86 & 0.49 & 1.16 & -9.15 & 2.90 & 249.68 & 262.80 & 4.09e+06 & 5.13e+03 & \multirow{2}{*}{not full} \\
 &$\pm$ 0.00 & $\pm$ 0.01 & $\pm$ 0.00 & $\pm$ 0.07 & $\pm$ 0.09 & $\pm$ 0.01 & $\pm$ 0.07 & $\pm$ 0.01 & $\pm$ 0.10 & $\pm$ 0.69 & $\pm$ 0.00 & $\pm$ 0.15 & $\pm$ 1.92e+05 & $\pm$ 2.25e+02 & \\
 & & & & & & & & & & & & & & &  \\
\multirow{2}{*}{7} &-1.06 & -3.66 & -0.02 & 253.81 & -0.31 & 3.81 & 3.36 & 4.26 & -1.58 & 0.57 & 253.17 & 254.45 & 3.82e+06 & 5.02e+03 & \multirow{2}{*}{RCW 19} \\
 &$\pm$ 0.03 & $\pm$ 0.08 & $\pm$ 0.02 & $\pm$ 0.06 & $\pm$ 0.31 & $\pm$ 0.09 & $\pm$ 0.08 & $\pm$ 0.12 & $\pm$ 0.54 & $\pm$ 0.31 & $\pm$ 0.08 & $\pm$ 0.06 & $\pm$ 8.58e+05 & $\pm$ 1.13e+03 & \\
 & & & & & & & & & & & & & & &  \\
\multirow{2}{*}{8} &-0.03 & -1.12 & 0.07 & 268.54 & 3.68 & 1.12 & 0.91 & 1.39 & 0.86 & 5.29 & 263.73 & 272.82 & 2.17e+06 & 3.15e+03 & \multirow{2}{*}{not full} \\
 &$\pm$ 0.00 & $\pm$ 0.01 & $\pm$ 0.01 & $\pm$ 0.22 & $\pm$ 0.26 & $\pm$ 0.01 & $\pm$ 0.03 & $\pm$ 0.01 & $\pm$ 0.95 & $\pm$ 0.03 & $\pm$ 0.44 & $\pm$ 0.55 & $\pm$ 2.85e+05 & $\pm$ 4.16e+02 & \\
 & & & & & & & & & & & & & & &  \\
\multirow{2}{*}{9} &0.17 & -1.89 & 0.01 & 275.03 & 0.34 & 1.90 & 1.66 & 2.09 & -0.73 & 1.45 & 273.81 & 276.37 & 1.45e+06 & 1.76e+03 & \multirow{2}{*}{} \\
 &$\pm$ 0.00 & $\pm$ 0.00 & $\pm$ 0.00 & $\pm$ 0.01 & $\pm$ 0.01 & $\pm$ 0.00 & $\pm$ 0.02 & $\pm$ 0.03 & $\pm$ 0.04 & $\pm$ 0.01 & $\pm$ 0.05 & $\pm$ 0.03 & $\pm$ 3.95e+04 & $\pm$ 4.14e+01 & \\
 & & & & & & & & & & & & & & &  \\
\multirow{2}{*}{10} &0.36 & -2.55 & 0.05 & 278.08 & 1.02 & 2.58 & 2.41 & 2.69 & 0.36 & 1.69 & 277.42 & 278.68 & 4.53e+05 & 5.18e+02 & \multirow{2}{*}{} \\
 &$\pm$ 0.00 & $\pm$ 0.01 & $\pm$ 0.00 & $\pm$ 0.01 & $\pm$ 0.02 & $\pm$ 0.01 & $\pm$ 0.02 & $\pm$ 0.02 & $\pm$ 0.07 & $\pm$ 0.05 & $\pm$ 0.01 & $\pm$ 0.01 & $\pm$ 3.47e+04 & $\pm$ 3.95e+01 & \\
 & & & & & & & & & & & & & & &  \\
\multirow{2}{*}{11} &-0.37 & -1.08 & 0.00 & 250.90 & 0.05 & 1.14 & 1.04 & 1.24 & -0.72 & 0.91 & 249.93 & 251.95 & 2.33e+05 & 1.94e+02 & \multirow{2}{*}{not full} \\
 &$\pm$ 0.00 & $\pm$ 0.00 & $\pm$ 0.00 & $\pm$ 0.01 & $\pm$ 0.01 & $\pm$ 0.00 & $\pm$ 0.00 & $\pm$ 0.01 & $\pm$ 0.01 & $\pm$ 0.01 & $\pm$ 0.00 & $\pm$ 0.07 & $\pm$ 6.23e+03 & $\pm$ 4.29e+00 & \\
 & & & & & & & & & & & & & & &  \\
\multirow{2}{*}{12} &-0.14 & -1.84 & -0.20 & 265.67 & -6.20 & 1.86 & 1.81 & 1.90 & -6.96 & -5.59 & 264.42 & 266.61 & 1.77e+05 & 1.94e+02 & \multirow{2}{*}{} \\
 &$\pm$ 0.00 & $\pm$ 0.03 & $\pm$ 0.00 & $\pm$ 0.10 & $\pm$ 0.02 & $\pm$ 0.03 & $\pm$ 0.01 & $\pm$ 0.07 & $\pm$ 0.02 & $\pm$ 0.10 & $\pm$ 0.37 & $\pm$ 0.23 & $\pm$ 1.10e+05 & $\pm$ 1.23e+02 & \\
 & & & & & & & & & & & & & & &  \\
\multirow{2}{*}{13} &-0.24 & -2.32 & -0.11 & 264.10 & -2.75 & 2.33 & 2.26 & 2.40 & -3.28 & -2.36 & 263.50 & 264.70 & 1.43e+05 & 1.68e+02 & \multirow{2}{*}{} \\
 &$\pm$ 0.00 & $\pm$ 0.03 & $\pm$ 0.00 & $\pm$ 0.06 & $\pm$ 0.04 & $\pm$ 0.03 & $\pm$ 0.04 & $\pm$ 0.02 & $\pm$ 0.12 & $\pm$ 0.12 & $\pm$ 0.12 & $\pm$ 0.06 & $\pm$ 3.62e+04 & $\pm$ 3.98e+01 & \\
 & & & & & & & & & & & & & & &  \\
\multirow{2}{*}{14} &-0.02 & -0.47 & -0.05 & 267.98 & -6.56 & 0.48 & 0.44 & 0.51 & -9.53 & -2.85 & 265.04 & 269.95 & 6.00e+04 & 4.07e+01 & \multirow{2}{*}{} \\
 &$\pm$ 0.00 & $\pm$ 0.02 & $\pm$ 0.00 & $\pm$ 0.37 & $\pm$ 0.28 & $\pm$ 0.02 & $\pm$ 0.05 & $\pm$ 0.02 & $\pm$ 0.79 & $\pm$ 1.10 & $\pm$ 0.58 & $\pm$ 1.83 & $\pm$ 1.77e+04 & $\pm$ 2.76e+01 & \\
 & & & & & & & & & & & & & & &  \\
\bottomrule
\end{tabular}

%% file: figures/cavitiesTable.tex
\begin{tabular}{cccccccccccccc}
\toprule
\multirow{2}{*}{ID} & X & Y & Z & l & b & d & mind & maxd & minb & maxb & minl & maxl & Volume \\
 & (kpc) & (kpc) & (kpc) & (deg) & (deg) & (kpc) & (kpc) & (kpc) & (deg) & (deg) & (deg) & (deg) & (pc$^3$) \\
\midrule
\multirow{2}{*}{1} &-0.22 & -3.17 & -0.17 & 266.11 & -3.15 & 3.18 & 1.19 & 4.66 & -7.43 & -0.81 & 261.37 & 269.46 & 1.05e+08 \\
 &$\pm$ 0.02 & $\pm$ 0.35 & $\pm$ 0.03 & $\pm$ 0.36 & $\pm$ 0.43 & $\pm$ 0.35 & $\pm$ 0.99 & $\pm$ 0.37 & $\pm$ 0.83 & $\pm$ 0.59 & $\pm$ 1.62 & $\pm$ 0.34 & $\pm$ 3.81e+07 \\
 &  &  &  &  &  &  &  &  &  &  &  &  & \\
\multirow{2}{*}{2} &0.31 & -3.68 & -0.19 & 274.88 & -2.88 & 3.69 & 3.23 & 4.01 & -4.44 & -1.40 & 272.58 & 277.01 & 1.79e+07 \\
 &$\pm$ 0.02 & $\pm$ 0.12 & $\pm$ 0.03 & $\pm$ 0.21 & $\pm$ 0.44 & $\pm$ 0.12 & $\pm$ 0.20 & $\pm$ 0.11 & $\pm$ 0.78 & $\pm$ 0.31 & $\pm$ 0.64 & $\pm$ 0.31 & $\pm$ 5.49e+06 \\
 &  &  &  &  &  &  &  &  &  &  &  &  & \\
\multirow{2}{*}{3} &-1.07 & -4.46 & -0.32 & 256.54 & -4.01 & 4.60 & 4.35 & 4.82 & -5.45 & -2.55 & 255.30 & 257.99 & 9.73e+06 \\
 &$\pm$ 0.10 & $\pm$ 0.35 & $\pm$ 0.02 & $\pm$ 0.96 & $\pm$ 0.11 & $\pm$ 0.35 & $\pm$ 0.82 & $\pm$ 0.12 & $\pm$ 0.59 & $\pm$ 0.51 & $\pm$ 1.75 & $\pm$ 0.91 & $\pm$ 2.18e+07 \\
 &  &  &  &  &  &  &  &  &  &  &  &  & \\
\multirow{2}{*}{4} &-0.42 & -3.51 & -0.17 & 263.18 & -2.76 & 3.54 & 3.03 & 3.97 & -3.78 & -1.53 & 262.31 & 264.03 & 7.11e+06 \\
 &$\pm$ 0.02 & $\pm$ 0.11 & $\pm$ 0.03 & $\pm$ 0.26 & $\pm$ 0.54 & $\pm$ 0.11 & $\pm$ 0.27 & $\pm$ 0.08 & $\pm$ 0.71 & $\pm$ 0.36 & $\pm$ 0.34 & $\pm$ 0.31 & $\pm$ 2.71e+06 \\
 &  &  &  &  &  &  &  &  &  &  &  &  & \\
\multirow{2}{*}{5} &0.19 & -2.91 & -0.03 & 273.75 & -0.61 & 2.91 & 1.99 & 3.75 & -1.50 & 0.33 & 272.09 & 274.87 & 5.95e+06 \\
 &$\pm$ 0.01 & $\pm$ 0.05 & $\pm$ 0.01 & $\pm$ 0.11 & $\pm$ 0.26 & $\pm$ 0.05 & $\pm$ 0.03 & $\pm$ 0.13 & $\pm$ 0.29 & $\pm$ 0.23 & $\pm$ 0.14 & $\pm$ 0.16 & $\pm$ 4.39e+05 \\
 &  &  &  &  &  &  &  &  &  &  &  &  & \\
\multirow{2}{*}{6} &-0.58 & -3.31 & -0.09 & 260.04 & -1.49 & 3.36 & 2.44 & 4.26 & -2.36 & -0.71 & 259.05 & 261.29 & 4.35e+06 \\
 &$\pm$ 0.02 & $\pm$ 0.09 & $\pm$ 0.00 & $\pm$ 0.26 & $\pm$ 0.03 & $\pm$ 0.09 & $\pm$ 0.22 & $\pm$ 0.35 & $\pm$ 0.19 & $\pm$ 0.17 & $\pm$ 0.42 & $\pm$ 0.05 & $\pm$ 1.33e+06 \\
 &  &  &  &  &  &  &  &  &  &  &  &  & \\
\multirow{2}{*}{7} &-0.29 & -1.54 & -0.17 & 259.32 & -6.12 & 1.58 & 1.43 & 1.77 & -7.23 & -4.80 & 258.23 & 260.76 & 8.47e+05 \\
 &$\pm$ 0.04 & $\pm$ 0.09 & $\pm$ 0.03 & $\pm$ 1.86 & $\pm$ 1.18 & $\pm$ 0.08 & $\pm$ 0.08 & $\pm$ 0.09 & $\pm$ 1.14 & $\pm$ 1.00 & $\pm$ 3.06 & $\pm$ 1.36 & $\pm$ 4.97e+05 \\
 &  &  &  &  &  &  &  &  &  &  &  &  & \\
\multirow{2}{*}{8} &0.41 & -3.04 & -0.03 & 277.62 & -0.53 & 3.07 & 2.50 & 3.60 & -0.81 & -0.24 & 277.30 & 278.01 & 5.95e+05 \\
 &$\pm$ 0.00 & $\pm$ 0.01 & $\pm$ 0.00 & $\pm$ 0.00 & $\pm$ 0.00 & $\pm$ 0.01 & $\pm$ 0.01 & $\pm$ 0.02 & $\pm$ 0.01 & $\pm$ 0.08 & $\pm$ 0.01 & $\pm$ 0.02 & $\pm$ 1.78e+04 \\
 &  &  &  &  &  &  &  &  &  &  &  &  & \\
\multirow{2}{*}{9} &0.02 & -3.40 & -0.19 & 270.42 & -3.22 & 3.40 & 3.24 & 3.56 & -3.35 & -3.06 & 270.32 & 270.68 & 8.70e+04 \\
 &$\pm$ 0.00 & $\pm$ 0.08 & $\pm$ 0.01 & $\pm$ 0.04 & $\pm$ 0.03 & $\pm$ 0.08 & $\pm$ 0.10 & $\pm$ 0.08 & $\pm$ 0.05 & $\pm$ 0.04 & $\pm$ 0.08 & $\pm$ 0.11 & $\pm$ 3.99e+04 \\
 &  &  &  &  &  &  &  &  &  &  &  &  & \\
\bottomrule
\end{tabular}